\documentclass{achemso}

\usepackage{textcomp}
\usepackage{amsmath}
\usepackage{amssymb}
\usepackage{graphicx}
\usepackage{esint}
\usepackage{sidecap}
\usepackage{color}
\usepackage{ulem}

\usepackage{caption} 
\usepackage{float}
\usepackage{geometry}
\usepackage{natbib}
\usepackage{setspace}
\usepackage{xkeyval}

\usepackage{subcaption}

\title{A physics based model of gate tunable metal-graphene contact resistance benchmarked against experimental data}

\author{Ferney A. Chaves}
\affiliation{Departament d'Enginyeria Electr\`{o}nica, Escola
d'Enginyeria, Universitat Aut\`{o}noma de Barcelona, Campus
UAB, 08193 Bellaterra, Spain.}

\email{ferneyalveiro.chaves@uab.cat}

\author{David Jim\'{e}nez}
\affiliation{Departament d'Enginyeria Electr\`{o}nica, Escola
d'Enginyeria, Universitat Aut\`{o}noma de Barcelona, Campus
UAB, 08193 Bellaterra, Spain.}

\author{Abhay A. Sagade}
\affiliation{Advanced Microelectronic Center Aachen (AMICA), AMO GmbH, Otto-Blumenthal-Strasse 25, 52074 Aachen, Germany}

\author{Wonjae Kim}
\affiliation{Aalto University, Department of Micro- and Nanosciences, Micronova, Tietotie 3, FI-02150 Espoo, Finland}

\author{Juha Riikonen}
\affiliation{Aalto University, Department of Micro- and Nanosciences, Micronova, Tietotie 3, FI-02150 Espoo, Finland}

\author{Harri Lipsanen}
\affiliation{Aalto University, Department of Micro- and Nanosciences, Micronova, Tietotie 3, FI-02150 Espoo, Finland}

\author{Daniel Neumaier}
\affiliation{Advanced Microelectronic Center Aachen (AMICA), AMO GmbH, Otto-Blumenthal-Strasse 25, 52074 Aachen, Germany}

\begin{document}

\begin{abstract}
The metal-graphene contact resistance is a technological bottleneck for the realization of viable graphene based electronics. We report a useful model to find the gate tunable components of this resistance determined by the sequential tunneling of carriers between the 3D-metal and 2D-graphene underneath followed by Klein tunneling to the graphene in the channel. This model quantifies the intrinsic factors that control that resistance, including the effect of unintended chemical doping. Our results agree with experimental results for several metals.

\end{abstract}

\maketitle


\begin{figure*}
\includegraphics[scale=0.5]{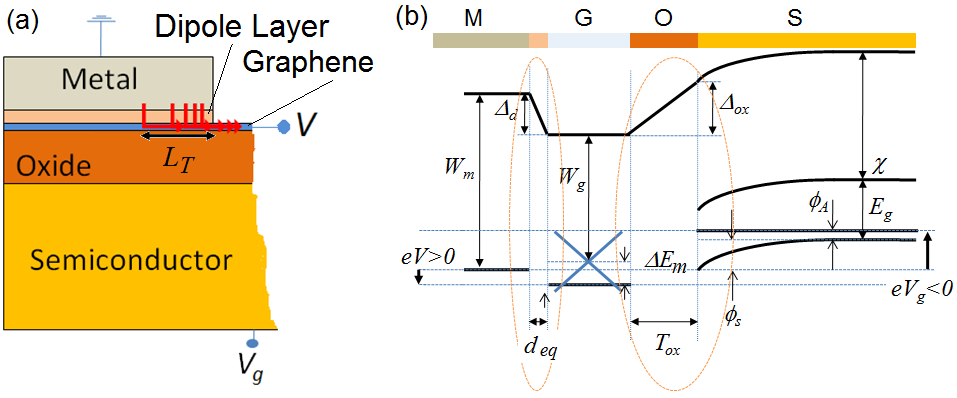}
\caption{ Sketch of the device considered in this work (a) and the band diagram of a MGOS heterostructure (b).
Red arrows suggest the current crowding effect near to the contact edge.}
\label{estructure}
\end{figure*}

\section{Introduction}

While graphene has emerged as a promising material for future electronic devices thanks to its unique electronic
properties, the metal-graphene contact resistance ($R_c$) remains a limiting factor for graphene-based electronic
devices \cite{Novoselov}. In particular, for high frequency electronics it is an issue very much influencing figures of merit like the maximum frequency of oscillation, the cutoff frequency, or the intrinsic gain \cite{Schwierz}.
Therefore it is neccesary to understand the intrinsic and extrinsic factors determining $R_c$, which displays a strong variation depending on the metal contact and fabrication procedure details\cite{Huard,Nagashio,Russo}. To gain understanding of these factors so that a better control of the contact's technology is feasible, a comprehensive physics based model of $R_c$ is an absolute requirement. One relevant model was already proposed by Xia \textit{et al.}\cite{Xia} to describe the transport in metal-graphene junctions as two sequential tunneling process from the metal to graphene over an effective transfer length ($L_T$), followed by injection to the graphene channel (see Fig. \ref{estructure}a). However, there is an important ingredient determining $R_{c}$ namely, the transmission from a 3D system (metal) to a 2D system (graphene), that has so far not been taken into account properly there in a physics basis. Evidence of the current crowding effect over $L_T$ has been reported by Sundaram \textit{et al.} using photocurrent spectroscopy \cite{Sundaram}.

In order to improve the current understanding, we have considered the issue of the carriers transmission between materials of different dimensionality. Specifically, we have developed a physics-based model where the first process is responsible for the resistance between the metal and the graphene underneath ($R_{mg}$) and the second process includes the resistance due to a potential step across the junction formed between the graphene under the metal and the graphene channel ($R_{gg}$). The total contact resistance is then the series combination of both contributions, $R_{c}=R_{mg}+R_{gg}$, accounting for any current crowding effect near the contact edge. The calculation of $R_{mg}$ and $R_{gg}$ are based on the Bardeen Transfer Hamiltonian (BTH) method \cite{Bardeen,Tersoff} and the Landauer approach \cite{Cayssol}, respectively. The BTH method allows us to get information about the matrix elements for the transition between 3D-metal and 2D-graphene states and combined with Fermi's golden rule, yields a compact expression for the specific contact resistivity $\rho_c$. On the other hand, the Landauer approach allows to get the conductance of carriers across the potential step between the graphene under the metal and the graphene in the channel, where the angular dependence transmission of Dirac fermions and the effective length of the potential have been taken into account. To model $R_c$ we have considered it as a building block of a FET device, so its value will strongly depend on the applied gate voltage. 

\section{Methods}
\subsection{Electrostatics}

In this paper we start with the graphene electrostatics. We considered a three terminal graphene FET (GFET) device controlled by a global back-gate voltage $(V_{g})$ as sketched in Fig. \ref{estructure}a, although it could be easily adapted to a device with both top- and back-gates as we will show later on. We split the electrostatic problem by considering two 1D heterostructures, namely, the Metal/Graphene/Oxide/Semiconductor (MGOS) and Graphene/Oxide/Semiconductor (GOS) heterostructures in the contact and channel regions, respectively. In Fig. \ref{estructure}b the corresponding band diagram of the MGOS heterostructure has been shown. In each of these regions we model the gate voltage dependence of
the graphene Fermi level relative to its Dirac energy, namely $\Delta E_{m}$ and $\Delta E_{g}$ for the graphene under the metal and graphene in the channel, respectively. The energy potential loops at the encircled interfaces in Fig. \ref{estructure}b together with the Gauss's law are considered, resulting in Eqs. \ref{eq:eltc}a-c. Because of the charge transfer between the metal and graphene, a dipole layer of size $d_{1}$ inside the equilibrium separation distance $d_{eq}$ is set up\cite{Khomyakov}. Also a difference $eV$ between the metal and the graphene Fermi level in the contact region, supplied by the drain terminal, has been assumed. The work-functions of the metal and graphene are $W_{m}$ and $W_{g}$, respectively.

\begin{subequations}
\label{eq:eltc}
\begin{align}
         &W_{m}=e\Delta_{d}+W_{g}+\Delta E_{m}-eV,\\
         &W_{g}+\Delta E_{m}=e\Delta_{ox}+W_{sc}-e\left(\phi_{s}+V_{g}-V\right),\\
         &Q_{sc}+Q_{m}+Q_{g}=0
\end{align}        

\end{subequations}

In Eq. \ref{eq:eltc}a, the term $\Delta_{d}$ is the potential drop in the dipole layer which can be expressed as $\Delta_{d}=\Delta_{tr}+\Delta_{ch}=-Q_{m}/C_{d}+\Delta_{ch}$, where $\Delta_{tr}$ corresponds to the charge transfer and $\Delta_{ch}$ to chemical potential interaction describing the short range interaction from the overlap of the metal and graphene wavefunctions \cite{Khomyakov,Chaves}. In Eq. \ref{eq:eltc}b, the back-gate voltage $V_{g}$ is referred to the source metal electrode potential, $W_{sc}=\chi+E_{g}-\phi_{A}$ is the semiconductor work-function and $\phi_{s}$ is the semiconductor surface potential. In Eq. \ref{eq:eltc}c, $Q_{m}=-C_d\Delta_{tr}$ describes the charge per unit area induced in the surface metal, $Q_{g}\approx 2e/(\pi\hbar^2v_f^2)\Delta E_m\vert\Delta E_m\vert+Q_0$ is the net charge sheet density within the graphene layer \cite{Feenstra} plus the charge density due to possible chemical doping\cite{ZhenHua} ($Q_0$)  and $Q_{sc}=C_{ox}\Delta_{ox}$ describes the charge per unit area induced in the semiconductor. Here, $C_{d}=\varepsilon_{0}/d_{1}$ and $C_{ox}=\varepsilon/T_{ox}$ describe the capacitive coupling to the metal and back gate, respectively. The value of $d_{1}$ can differ from the equilibrium distance $d_{eq}$ ($\sim0.3$nm) due to the spatial extension of the carbon $p_{z}$ and metal $d$ orbitals. The value of $\Delta_{ch}$ strongly depends on the separation distance $d_{eq}$ and it becomes negligible for $d_{eq}\gtrsim4$nm \cite{Khomyakov}. Combining Eqs. \ref{eq:eltc} and assuming that $\phi_s$ saturates at strong inversion and acummulation, we get a simple quadratic equation for $\Delta E_{m}$: 

\begin{equation}
a\Delta E_m\vert\Delta E_m\vert+\left(C{}_{ox}+C_{d}\right)\Delta E_{m}+eC_{ox}\left(V_{g}-V_{D}\right)=0\label{eq:DE}
\end{equation}
where $a=e^2/\pi\hbar^2 v_f^2$, with $v_f (\sim1\times10^8$ cm/s)  the Fermi velocity,  and 

\begin{equation}
eV_{D}= \dfrac{C_{d}}{C_{ox}} \left(W_{m}-W_{g}+eV-e\Delta_{ch}\right)
+\left(\chi+eV+e\phi_{A}-W_{g}-\dfrac{Q_0}{C_{ox}}\right)
\label{eq:VgD}
\end{equation}
represents the Dirac gate voltage required to achieve $\Delta E_{m}=0$ and defines the back-gate voltage value for which $\rho_c$ and the resistance $R_{mg}$ become maximum, as we will see later.

Because the dipole layer has been modeled as an insulator, the channel region electrostatics under the influence of both top- and back-gates can be described in a similar way as presented in Eqs. \ref{eq:eltc}, so the Fermi level shift in the channel ($\Delta E_g$) can be obtained from:

\begin{equation}
a\Delta E_g\vert\Delta E_g\vert+\left(C_b+C_t\right)\Delta E_g+eC_b\left(V_b-V_{gD}\right)=0,
\label{eq:DEg}
\end{equation}
with
\begin{equation}
eV_{gD}= \dfrac{C_t}{C_b} \left(W_m-W_g-eV_t\right)+\left(\chi+e\phi_{A}-W_{g}-\dfrac{Q_0}{C_{b}}\right).
\label{eq:VggD}
\end{equation}
 
In the last equations, $C_{b(t)}$ and $V_{b(t)}$ are the  back (top)- capacitance and gate voltage, respectively. The new Dirac voltage $V_{gD}$ must be understood as the back-gate voltage needed to achieve $\Delta E_g=0$ at a fixed top-gate voltage $V_t$. When there is only a back-gate, like in our experimental devices, we can get $\Delta E_{g}$ for the GOS structure simply setting $C_{t}\rightarrow0$.

\begin{figure}
\includegraphics[scale=0.30]{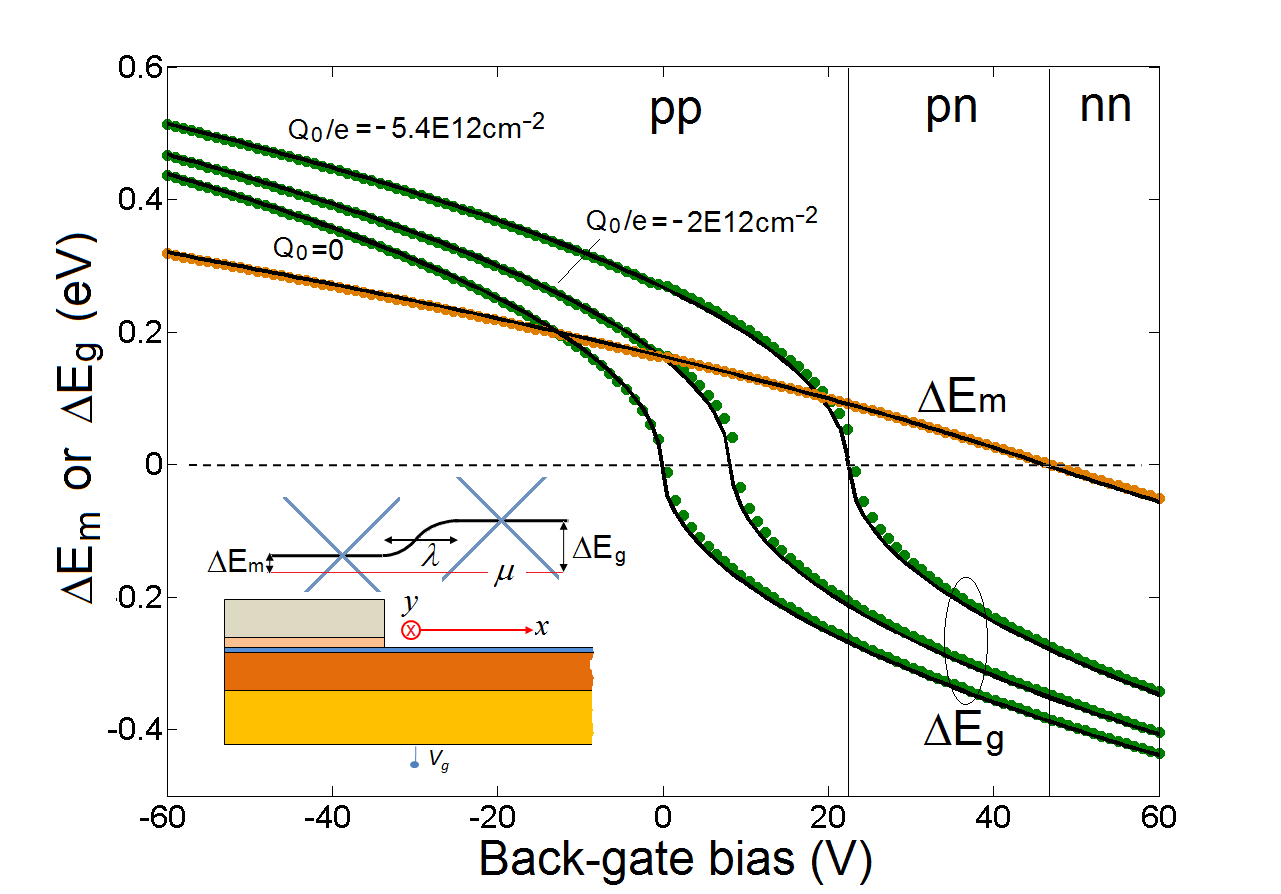}
\caption{Graphene Fermi level shifts with respect to the Dirac point for different values of $Q_0$ using Pd as metallic contact. Solid lines: numerical solution of Eq. (1) and symbols: solution of Eq. (2). The inset shows the potential step between the graphene underneath the metal and the graphene in the channel, with effective length $\lambda$.}
\label{shiftfermis}
\end{figure}

With the electrostatic model given by the above equations, the key quantities $\Delta E_{m}$ and $\Delta E_{g}$ could be determined, which in turn are needed to calculate the contact resistance. Fig. \ref{shiftfermis} shows both $\Delta E_{m}$ and $\Delta E_{g}$ at equilibrium $(V\rightarrow0)$ as a function of the back-gate bias using Palladium (Pd) as metal electrode and SiO$_{2}$ as oxide with thickness $T_{ox}=90$nm. This can be done either by solving Eqs. \ref{eq:eltc} or the simplified Eq. \ref{eq:DE}, with a very little difference between them. Different kinds of junctions may build-up depending on the back-gate bias, namely pp-type, pn-type, and nn-type. Here, we have assumed that $Q_0$ is only affecting the graphene channel, and not the graphene underneath the metal. The impact of $Q_0$ in determining the crossing of $\Delta E_g$ with zero, can be seen in the figure. To capture the transition between the pp-type and pn-type junction, which was observed at $V_g=V_{gD}\sim 23$V\cite{Xia}, the parameter $Q_0/e$ was set to $-5.4\times10^{12}$cm$^{-2}$. Next transistion produced between the pn-type and nn-type junction was captured by our model at $V_g=V_D\sim46$V, in accordance with the experiment of Xia \textit{et al}. The electrical parameters that we have assumed  for all the simulations presented in this work are shown in Table I. 
Because of charge transfer between the graphene underneath the metal and the graphene in the channel, a potential step of effective length $\lambda$ arises at the contact edge. An sketch of that potential step is illustrated in Fig. \ref{shiftfermis}. Once we get the electrostatic model, we are now ready to discuss how to model the contact resistance.

\subsection{Resistance $R_{mg}$ and resistivity}

The procedure to model $R_{mg}$ is based on the Transmission Line Method \cite{Schroder,Leonard,Reeves}, which in turn requires determination of $\rho_{c}$, namely:

\begin{equation}
R_{mg}(\Delta E_m)=\sqrt{\rho_{c}R_{sh}^{m}}\coth\left(L_{c}/L_{T}\right)/W,\label{eq:Rmg}
\end{equation}

where $\rho_{c}(\Delta E_m)=\left(dJ/dV\right)^{-1}\mid_{V=0}$ represents the specific contact resistivity, $R_{sh}^{m}$ (250$\Omega/\square$ in this work) is the graphene sheet resistance under the metal, $L_{T}=\sqrt{\rho_{c}/R_{sh}^{m}}$ is the characteristic length over which current injection occurs between the metal and the graphene layer (transfer length), and $L_{c}\,(W)$ is the length (width) of the contact. Here, $\rho_{c}$ is calculated by means of the BTH method, which allows us to split the metal-graphene system into separate metal and graphene subsystems with known Hamiltonians. In the framework of the BTH method, the probability of elastic tunneling is calculated using Fermi's golden rule. This gives a quantitative estimate of the coupling between the metal and graphene states, so it is possible to get an analytical formula with key parameters for $\rho_{c}$ as a function of $\Delta E_{m}$. In the Supplementary data we show how to calculate $\rho_{c}$ from the tunneling current density $J$ using the BTH approach. The resulting compact analytical expression for $\rho_{c}$ as a function of $\Delta E_{m}$ under the metal at $V=0$, for a given temperature $T$ can be written as:

\begin{equation}
\rho_{c}(\Delta E_{m})=\frac{\pi \gamma \hbar^3 v_f^2\exp(2\kappa d_{eq})}{16e^2}
\frac{\left(E_{\parallel}-\Delta E_{m}+\gamma^2E_{\kappa}\right)}{\sqrt{E_{\kappa}^3(E_{\parallel}-\Delta E_{m})}}
\frac{1}{2kT\,\ln\left(\exp(\Delta E_{m}/kT)+1\right)-\Delta E_{m}}\label{eq:rho}
\end{equation}

\begin{figure}
\includegraphics[scale=0.30]{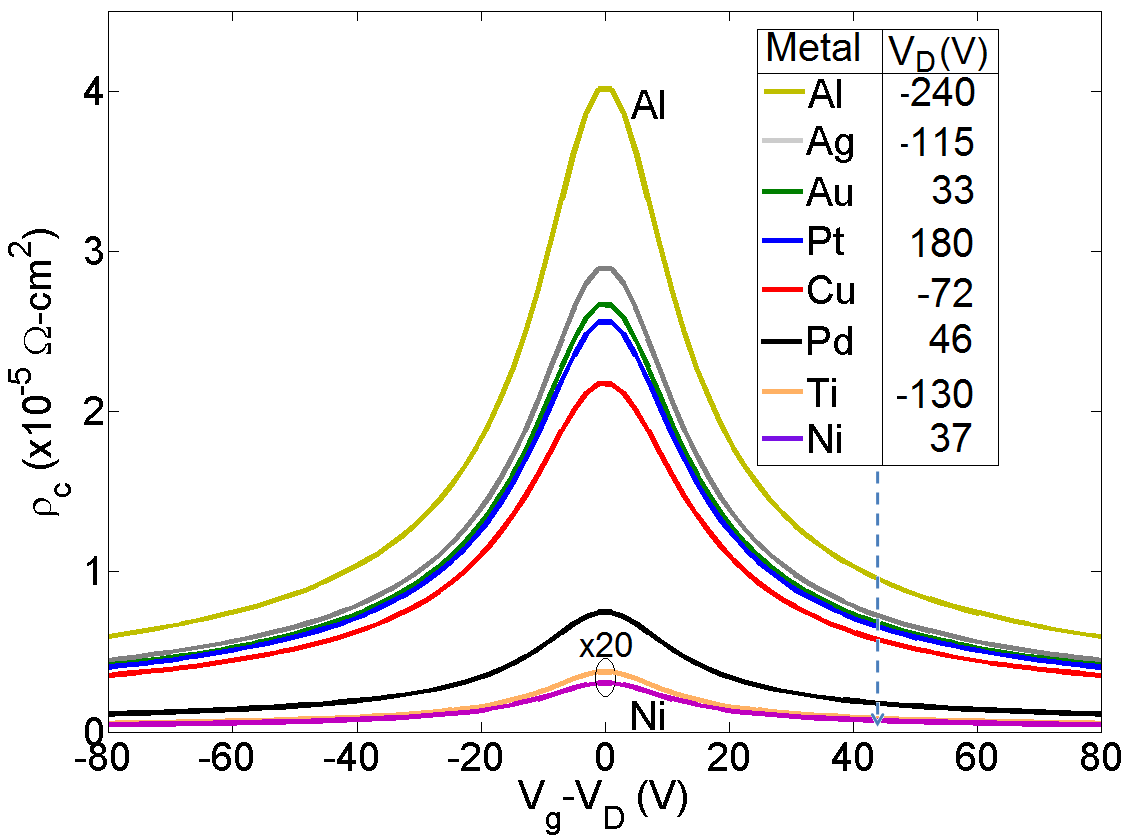}
\caption{Specific contact resistivity at room temperature for different metal electrodes centered
at $V_{g}=V_{D}$. The contact resistivities of Ni and Ti were multiplied by 20 in this plot.}
\label{fig:rho}
\end{figure}

where $\gamma=m/m_0$, with $m$ and $m_{0}$ the effective electron mass in the metal and dipole layer, respectively. The factor $\kappa$ is the electron decay constant in the dipole layer and has the form $\sqrt{2m\phi/\hbar^{2}+k_{\parallel}^{2}}\ $\cite{Feenstra}, where $\phi\sim W_m$ has been taken as the barrier height and $k_{\parallel}$ is the parallel momentum at the K or K$^\prime$ points (i.e., $4\pi/3a$). For a typical metal the work-function is $W_m\sim5$eV so $\kappa\sim20$nm$^{-1}$. As a consequence $E_\parallel=\hbar^{2}k_\parallel^{2}/2m\sim11$eV and $E_{\kappa}=\hbar^{2}\kappa^{2}/2m\sim16$eV. From Eq. \ref{eq:rho} we can infer that the maximum value of $\rho_{c}$ depends exponentially on the equilibrium separation distance $d_{eq}$ and that the maximum resistivity is located at $V_{D}$ ($\Delta E_m=0$). Fig. \ref{fig:rho} shows $\rho_{c}$ at $T=300$K as a function of the back-gate bias overdrive ($V_{g}-V_{D}$) considering different metals. After sorting the metals by their peak contact resistivity, it appears that $d_{eq}$ is the main factor controlling it, being the Ni contact the best option, followed by Ti. Here we have assumed SiO$_{2}$ as the insulator with $T_{ox}=90$nm and equal effective masses for every metal. According to Table I, the Ni-graphene (Al-graphene) contact has the smallest (largest) equilibrium distance $d_{eq}$ of the metals here represented, giving rise to the smallest (largest) value of $\rho_{c}$ at $V_{D}$. The values of $\rho_{c}$ predicted from Eq. \ref{eq:rho} are consistent with experimental results reported by Nagashio and Berdebes \cite{Nagashio2,Berdebes} for Ni, Ti and Pd. Although the Ni happens to be the best option to get the lowest $R_c$, other effects that contribute to the lateral resistance must be considered. As a matter of fact, $R_c$ for Ni can become comparable to that of Pd, as we will show later. Our model predicts how $\rho_c$ depends on factors like the workfunction difference, the equilbrium distance, the chemical interaction potential, the gate capacitance and the temperature.

\begin{table}
\resizebox{6cm}{!} {
\begin{tabular}{|c|c|c|c|c|}
\hline 
Metal & $W_{m}(eV)$ & $d_{eq}(\mathring{A})$ & $\Delta_{ch}(eV)$  \tabularnewline
\hline 
\hline 
Ni & 5.47 & 2.05 & 0.8$^*$ \tabularnewline
\hline 
Ti & 4.65 & 2.10 & 0.9$^*$ \tabularnewline
\hline 
Pd & 5.67 & 3.00 & 0.90    \tabularnewline
\hline 
Cu & 5.22 & 3.26 & 0.99    \tabularnewline
\hline 
Pt & 6.13 & 3.30 & 0.93    \tabularnewline
\hline 
Au & 5.54 & 3.31 & 0.91    \tabularnewline
\hline 
Ag & 4.92 & 3.33 & 0.88    \tabularnewline
\hline 
Al & 4.22 & 3.41 & 0.77    \tabularnewline
\hline 
\end{tabular}
}
\caption{
Electrical parameters for selected metal electrodes. They were extracted from previous reports\cite{Xia, Khomyakov}, except the quantities marked with $"*"$ which were considered as fitting parameters to match the experimental results.
}
\end{table}

\subsection{Resistance $R_{gg}$}

Next, we model the lateral contact resistance $R_{gg}$ across a potential step with effective length $\lambda$ (see inset of Fig. \ref{shiftfermis}) relying on the Landauer approach. The potential along the transport direction $x$ can be described by a simple space-dependent Fermi level shift\cite{Cayssol}:

\begin{equation}
\Delta E(x)=\Delta E_{m}+\frac{\Delta E_{g}-\Delta E_{m}}{\exp\left(-x/\lambda\right)+1},\label{eq:step}
\end{equation}

where we have considered that the metal electrode cover the left half-plane ($x<0$). The type ($n$ or $p$) and density of carriers in both left and right half-planes are tuned by the back-gate. The important quantity to be determined is the reflection probability of Dirac fermions across the potential step, which has been derived by Cayssol \textit{et al.}\cite{Cayssol}, namely:

\begin{equation}
R_{step}=\frac{\sinh(\pi\lambda\kappa^{+-})\sinh(\pi\lambda\kappa^{-+})}{\sinh(\pi\lambda\kappa^{++})\sinh(\pi\lambda\kappa^{--})}\label{eq:Reflexion}
\end{equation}

where the momenta $\kappa^{\rho\sigma}=\left(\Delta E_{g}-\Delta E_{m}\right)/\hbar v_{f}+\rho k_{x}^{(g)}+\sigma k_{x}^{(m)},$ with $\rho,\sigma=\pm1.$ The longitudinal momentum $k_{x}$ is related to the transversal momentum $k_{y}$ by the phytagorean relationship 

\begin{equation}
k_{x}^{(i)}=\mathrm{sgn}\left(\Delta E_{i}\right)\sqrt{\left(\Delta E_{i}/\hbar v_{f}\right)^{2}-k_{y}^{2}},\,\,\,\,\,\, i=m,g.\label{eq:kx}
\end{equation}

where the positive (negative) sign indicates that the doping is $p$ $(n)$ type.

\begin{figure}
\includegraphics[scale=0.18]{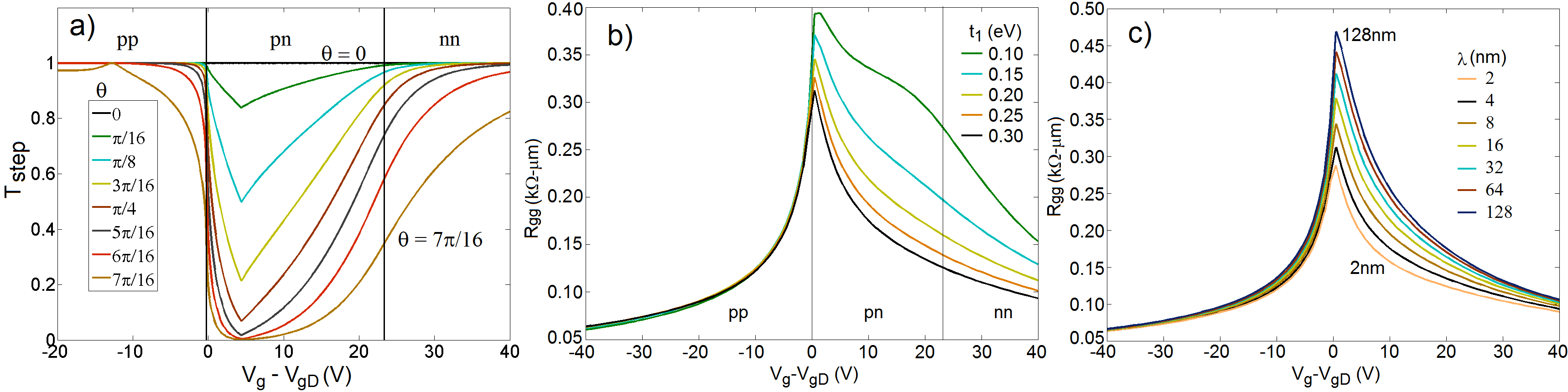}
\caption{(a) Transmission probability through a potential step with $\lambda=4$nm for different angles of incidence. Lateral contact resistance $R_{gg}$ for different values of $t_1$ and $t_2\sim100$meV (b) and different values of $\lambda$ (c). We have considered here Pd as the metal electrode.
}
\label{fig:3en1}
\end{figure}

By means of the Landauer formula the conductance can be obtained from: 

\begin{equation}
R_{gg}^{-1}(\Delta E_m,\Delta E_g)=\frac{2e^{2}}{h}\frac{W}{\pi}\int_{-k_{F}}^{k_{F}}T_{step}dk_{y}\label{eq:landauer}
\end{equation}

where $T_{step}=1-R_{step}$ is the transmission probability and $k_{F}=\min\left(\left|\Delta E_{m}\right|,\left|\Delta E_{g}\right|\right)/\hbar v_{f}$. Fig. \ref{fig:3en1}a shows the transmission probability of the Dirac fermions across the potential step as a function of $V_g$ for different incidence angles assuming Pd as the metal. In particular, it indicates the absence of backscattering at normal incidence ($k_{x}^{(m,g)}=k_{F}^{(m,g)}$ or $\theta=0$), because of the orthogonality of incoming and reflected spinor states. In contrast, the transmission of the bipolar contacts (case $pn$) tends toward zero for incident carriers when $\theta\rightarrow\pi/2$.

So far we have not considered the effect of the drain bias ($V_d$) in defining the contact resistance at the drain side ($R_{cd}$). However, for Radio-Frequency (RF) applications, $V_d$ is usually placed in the saturation region, so its value could be high as compared with $V_g$. In such a case the drain and source contact resistances can be very different. Our model for $R_c$ is still valid and useful to determine $R_{cd}$ in this situation. For this purpose it would be needed to evaluate $R_c$ at the effective gate voltage $V_{g,eff}=V_g-V_d$ instead of $V_g$, namely $R_{cd}=R_c(V_{g,eff})$.

\section{Results and discussion}

Until now, in the description of our model, we have not taken into account any broadening to the graphene states in the $R_c$ model. To get a more realistic model, an effective broadening describing the coupling between the metal and the quasi-bounded graphene states underneath and/or the spatial variations of the graphene-metal distance in the contact surface \cite{Khomyakov}, must be taken into account. This effect can be considered upon application of a Gaussian function $G_1$ of width $t_1$ (broadening energy). In addition, we have included the random disorder potential in the graphene channel using a Gaussian function $G_2$ of width $t_{2}=\hbar v_f\sqrt{2\pi n_0}$ where $n_0$ is the minimum sheet carrier concentration. Then, the two components of $R_c$ have to be recalculated as shown in Eqs. S16-17 of the Supplementary data.

In Fig. \ref{fig:3en1}b we show the effect of $t_1$ on 
$R_{gg}$ when it varies from $100$ to $300$ meV with $t_2\sim 100$meV ($n_0=5\times10^{11}$cm$^{-2}$). $R_{gg}$ exhibits a main
peak corresponding to the minimum DOS in the channel
($\Delta E_g=0$ or equivalently $V_g-V_{gD}=0$) and another secondary peak corresponding to the minimum DOS in the graphene under the metal ($\Delta E_m=0$ or equivalently $V_g-V_{gD}\sim23$ V).
According to the experimental data reported by Xia \textit{et al.}\cite{Xia} for Pd as metal electrode, the latter peak does not appear in the $R_c$ curve, suggesting a large $t_1$ ($>300$ meV) value as reflected in Fig. \ref{fig:3en1}b.

As a complementary information, the dependence of $R_{gg}$ on the effective length $\lambda$ of the potential step between the metal-doped graphene and the gate-controlled graphene channel is presented in Fig. \ref{fig:3en1}c. For unipolar juntions, $R_{gg}$ is almost independent of $\lambda$ while for the bipolar $pn$ junction it moderately increases as $\lambda$ changes from 2 to 128 nm.

After presentation of the $R_c$ model, next is benchmarking it against experimental measurements in graphene FETs using the transfer length method (TLM) for metal electrodes such as Palladium (Pd), Nickel (Ni) and Titanium (Ti) as shown below.

In Fig. \ref{fig:Rc} we have plotted the data reported by Xia \textit{et al.} considering Pd as metal electrode. Here the graphene sheet was transferred to SiO$_2$ of 90 nm thickness. Our model reveals that $R_{mg}$ and $R_{gg}$ play a similar role. The absence of a peak in the experimental $R_{c}$ data at $V_{g}\sim 46$V suggests a large value of $t_{1}$, as it has previously been discussed. To match the experimental data we have assumed $Q_0/e=-5.4\times10^{12}$cm$^{-2}$, $t_1=300$meV, $t_2=100$meV and $\lambda=100$nm. Interestingly we capture the correct value of the Dirac voltage at $V_{gD}\sim23$V and the moderate asymmetry between the left and right branches: being $R_c$ lower for the left branch because of the much better carrier transmission of the unipolar pp junction as compared with the bipolar pn junction (see Fig. \ref{fig:3en1}a).

\begin{figure}
\includegraphics[scale=0.30]{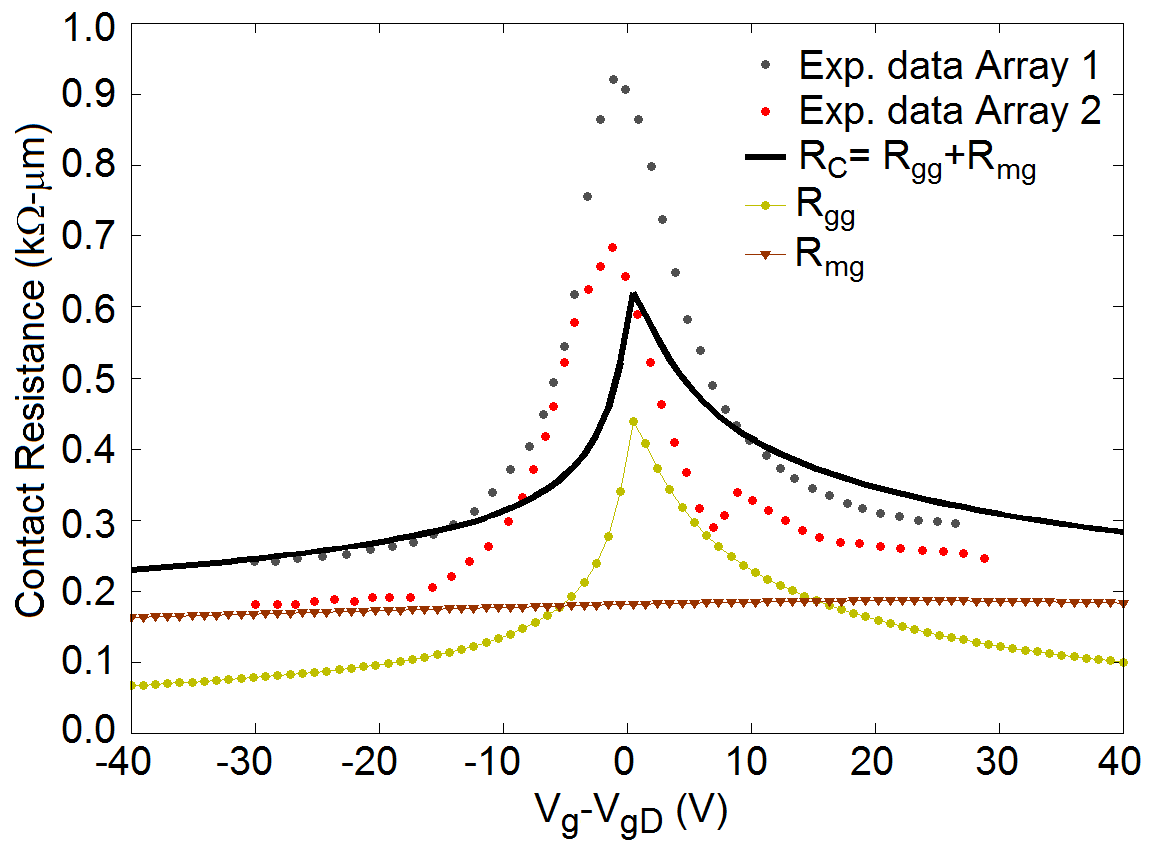}
\caption{Contact resistance and breakdown into its components as a function of the  back-gate bias overdrive for Pd as
metal electrode, where $\lambda$ has been assumed equal to 100nm.} 
\label{fig:Rc}
\end{figure}

\begin{figure}
\includegraphics[scale=0.30]{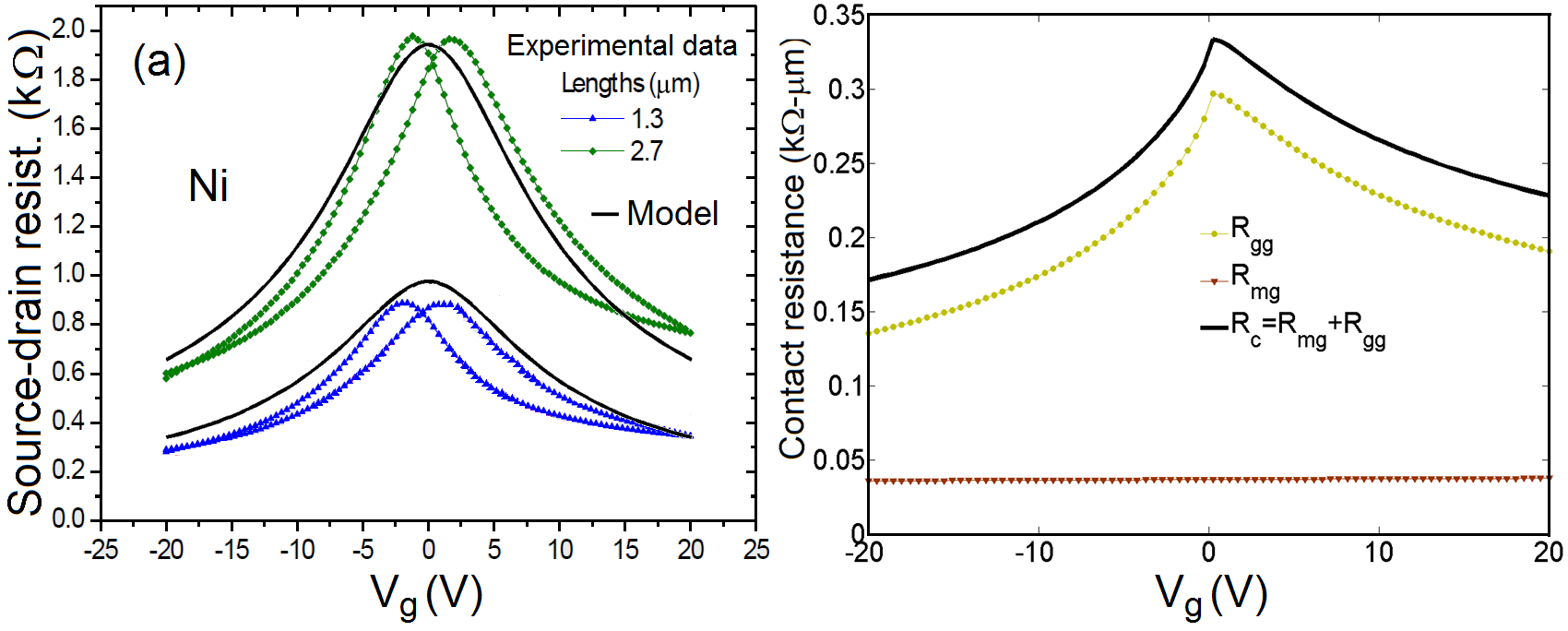}
\caption{(a) Comparison of the experimental (dotted lines) and simulated (black solid lines) total resistance between source and drain for a Ni contacted graphene FET. (b) Predicted $R_c$ and corresponding breakdown into $R_{mg}$ and $R_{gg}$ components.} 
\label{fig:RsdNi}
\end{figure}

Next, we compare with experimental data of GFETs with Ni as metal electrode (Fig. \ref{fig:RsdNi}). In this case, the back-gated graphene transistors have been fabricated by photolithography on Si wafers covered by 300 nm of thermal oxide. Graphene grown by chemical vapor deposition (supplier Bluestone Global Tech) was transferred by the standard PMMA method \cite{Cai} to the substrate and patterned using oxygen plasma. Nickel-contacts have been fabricated using sputter deposition and lift-off technique. The distance between source and drain contacts was 0.6, 0.9, 1.3, 1.7 and 2.7 $\mu$m for different devices on the chip to allow extraction of $R_c$ by TLM. The channel width was 10$\mu$m. Finally the devices have been encapsulated by 85 nm of Al$_2$O$_3$ deposited by atomic layer deposition. After some electrical measurements, we report in Fig. \ref{fig:RsdNi}a the comparison between the experimental data and the usual model of the source to drain resistance $R_T$ given by\cite{KimS}:  
\begin{equation}
R_T=\dfrac{R_{sh}^{ch}}{W}L_{ch}+2R_c.
\label{Eq:RT}
\end{equation}
Here, the channel sheet resistance $R_{sh}^{ch}$ has been modeled as $R_{sh}^{ch}=\left[ \mu e\sqrt{n_0^2+n(V_g)^2}\right]^{-1} $, with $\mu =1793$cm$^2$V$^{-1}$s$^{-1}$ and $n_0=5\times10^{11}$cm$^{-2}$ which were extracted from the experiment, and $n\propto\Delta E_g^2$ is the charge sheet concentration in the graphene channel region. In this case we have assumed a possible doping concentration in the graphene channel of $Q_0/e=-2\times10^{10}$cm$^{-2}$ in order to capture the position of the Dirac voltage. Details of the electrostatic behavior of the Ni-graphene contact can be found in the Supplementary data. For the quasi-static measurements of resistance shown in Fig. \ref{fig:RsdNi}a hysteretic behavior is observed, which is typical for graphene FETs. This hysteretic behavior occurs mainly because of charge traps generated by adsorbates, typically O$_2$/H$_2$O redox couples, at the graphene/dielectric interface \cite{Xu,Lee}. This effect has not been considered in this model.
Regarding the contact resistance (Fig. \ref{fig:RsdNi}b), our model gives values between 150 and 350 $\Omega$-$\mu$m, which are consistent with the experimental values extracted by TLM for the gate voltages $V_g=$ -20, 0 and 20$V$: $R_{c}\sim$ 220, 400 and 220 $\Omega$-$\mu$m  with correlation coefficient $R^2=0.9894, 0.9740$ and $0.9754$, respectively. The values of $\lambda$ and $t_1$ were determined to be around  4nm and 300meV, respectively, to get $R_c$ values in that range.  It is worth metioning that $R_{gg}$ is the dominant part of $R_c$, which is in contrast with the Pd contact case analyzed before, where $R_{gg}$ and $R_{mg}$ played a similar role.

\begin{figure}
\includegraphics[scale=0.28]{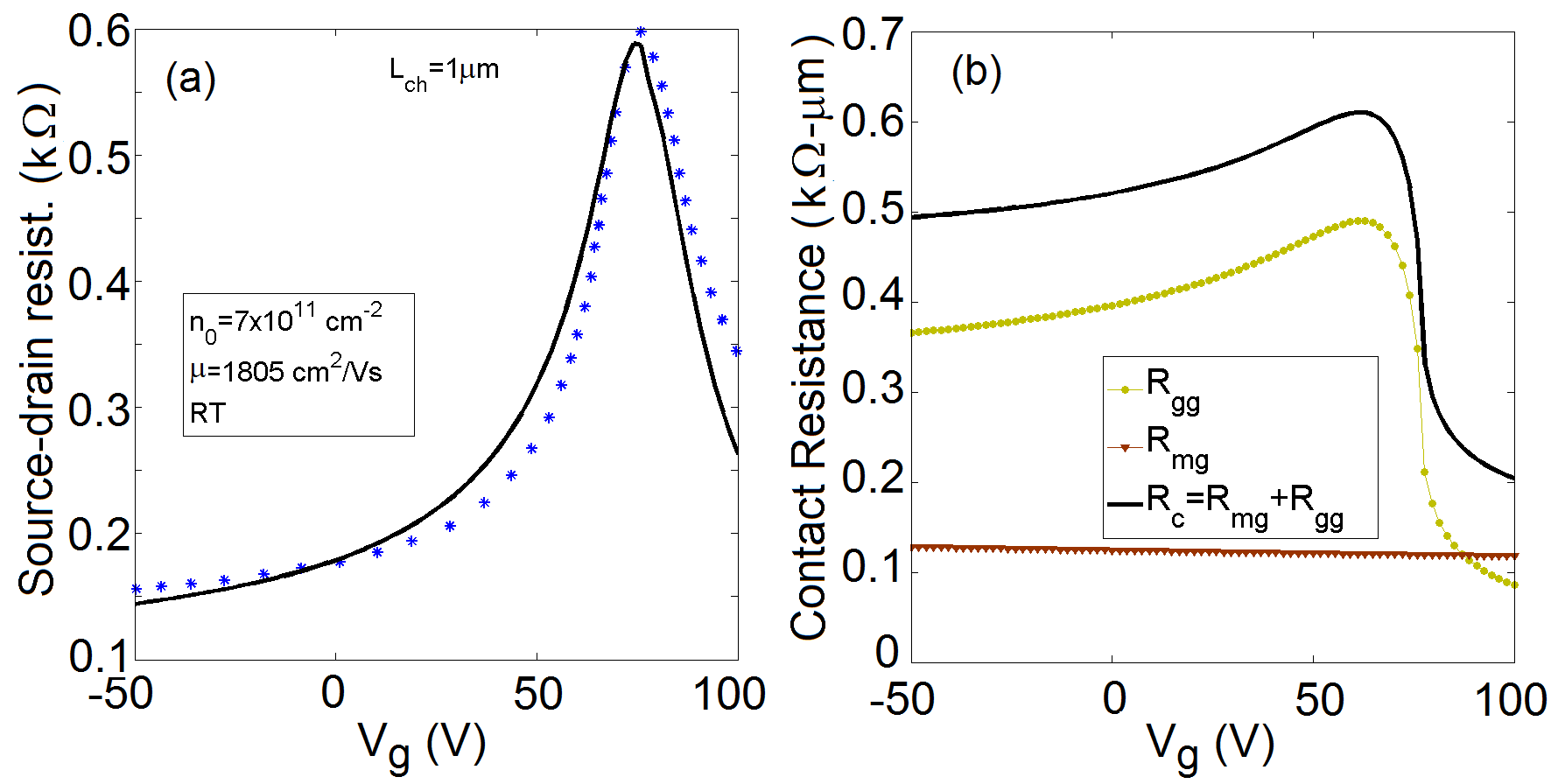}
\caption{(a) Comparison of the experimental (dotted lines) and simulated (black solid lines) total resistance between source and drain for a Ti contacted graphene FET. (b) Predicted $R_c$ and corresponding breakdown into $R_{mg}$ and $R_{gg}$ components.}  
\label{fig:RsdTi}
\end{figure}

A third comparison was carried out for GFETs  with Ti as metal electrode with geometrical parameters: $L_{ch}=1\mu$m, $W=10\mu$m and $T_{ox}=360$nm ($SiO_2$). Here graphene synthesized by photo-thermal CVD on copper was used to fabricate GFETs \cite{Riikonen, Kim}.  Regarding the source-drain resistance, the experimental data are shown in Fig. \ref{fig:RsdTi}a together with the model prediction on $R_T$. Similarly to the Ni case, we have considered the following electrical parameters: $\mu=1805$cm$^2$/Vs, $n_0=7\times10^{11}$cm$^{-2}$ as extracted from the experimental data. A chemical doping $Q_0/e=-4.6\times10^{12}$cm$^{-2}$ was fed in the model to get the position of the Dirac voltage $V_{gD}$ around 75V in accordance with the observation. Details of the electrostatic behavior of the Ti-graphene contact can be found in the Supplementary data. Using them together with the parameters given in Table I, our model results in the contact resistance shown in Fig. \ref{fig:RsdTi}b. The calculated $R_c$ agrees well with the extracted values from TLM at gate voltages $V_g= -75, 15, 45$ and $75$V:  $R_c \sim 500, 500,  400$ and $600\Omega$-$\mu$m with correlation coefficient $R^2=0.9932, 0.9915, 0.9986$ and $0.9997$, respectively. The values of $\lambda$ and $t_1$ were determined to be around  50nm and 300meV, respectively, to get $R_c$ values in the mentioned range.  Unlike Pd and Ni, $R_c$ in Ti-graphene contact exhibits a huge asymmetry between left and right branches, being $R_c$ lower for the right branch. This asymmetry qualitatively agrees with  measurements carried out for Ti and reported by Xia \textit{et al.}\cite{Xia}.           

\section{Conclusions}

In summary, we have developed a model of the gate tunable metal-graphene contact resistance. First of all we have modeled the behavior of the shift Fermi level $\Delta E$ in both the graphene underneath the metal and graphene in the channel. $\Delta E$ becomes zero under the metal at gate voltage named as $V_D$ which is controlled by intrinsic electrical parameters like the work function, the capacitive coupling between the metal and the gate and the value of the chemical interaction. In the channel region $\Delta E$ is zero at gate voltage $V_{gD}$ which is strongly determined by the unintended chemical doping. Once we get $\Delta E$ in each region, we use a combination of the BTH and the Landauer formula to independently determine the contribution of each $R_c$ component, namely BTH to determine the resistance between metal and the graphene underneath ($R_{mg}$) and Landauer formula for the resistance between graphene under the metal and the graphene in the channel ($R_{gg}$). Using BTH we have found a simple analytical expression for the specific contact resistivity $\rho_c$ which elucidates its dependence with the metal-graphene equilibrium distance. Specifically, among the metals considered here, Ni and Ti exhibit the smallest value of $\rho_c$ at their respectives Dirac voltages $V_D$. However, given the voltage dependence of $\rho_c$ and the different $V_D$ value displayed by each metal metal, Cu or Pd could show even a smaller $\rho_c$ than that for Ni or Ti depending on the applied gate voltage. The calculation of $\rho_c$ is key to get $R_{mg}$  by means of Transmission Line Method. This resistance shows a peak at $V_g=V_D$. On the other hand, the lateral resistance or $R_{gg}$, in principle, exhibits two peaks. One of them at $V_{gD}$ and another at $V_D$. However when a broadening of the graphene states under the metal ($t_1$ in this work) is considered, the latter peak could disappear. We have also found that $R_{gg}$ is sensitive to the effective length ($\lambda$) of the junction potential step, specially when a bipolar pn junction builds up.  Depending on the metal electrode and the chemical doping of the graphene channel the two components of $R_c$ could be either similar in magnitude or of very different order. In particular for Pd those two components compete, but for Ni and Ti the lateral resistance is the dominant component.
  
Our model is in agreement with experimental data for several metals under test. In particular, we have benchmarked the model against experiments using Pd, Ti, and Ni. The proposed model unveils the interplay between different intrinsic and extrinsic factors in determining the contact resistance of graphene-based electronic devices, which should be useful for its optimization.

\section{Acknowledgements}

We acknowledge support from SAMSUNG within the Global Research Outreach Program.
The research leading to these results has received funding from Ministerio
of Econom\'{i}a y Competitividad of Spain under the project TEC2012-31330
 and from the European Union Seventh Framework Programme
under grant agreement n\textdegree{}604391 Graphene Flagship.

\section{Supporting Information}
\subsection{Calculation of the specific contact resistivity}
In this section we derive the analytical expression for the specific contact resistivity of the Metal-Graphene junction given by Eq. (5) of the main text, relying on the BTH approach. The starting point is the expression for the tunneling current 

\begin{equation}
I=g_{S}g_{V}\sum_{g,m}\left\{ \Gamma_{gm}f_{g}(E_{g})[1-f_{m}(E_{m})]-\Gamma_{mg}f_{m}(E_{m})[1-f_{g}(E_{g})]\right\} \label{eq:current}
\end{equation}

where both the subscripts $g$ and $m$ label the states in the graphene and metal electrodes with energies $E_{g}$ and $E_{m}$, respectively, $g_{S}$ is the electron spin degeneracy, $g_{V}$ is the valley degeneracy, and $\Gamma_{gm}$ and $\Gamma_{mg}$ refer to the tunneling rates for electrons moving from $g\rightarrow m$ and $m\rightarrow g$, respectively. Finally, $f_{g}$ and $f_{m}$ are the Fermi occupation factors for the electrons.The tunneling rates
are given by the Fermi's golden rule as

\begin{equation}
\Gamma_{gm}=\frac{2\pi}{\hbar}|M_{gm}|^{2}\delta(E_{g}-E_{m})=\Gamma_{mg},\label{eq:gamma}
\end{equation}

where

\begin{equation}
M_{gm}=\frac{\hbar^{2}}{2m_{0}}\iint\left(\Psi_{g}^{*}\frac{d\Psi_{m}}{dz}-\Psi_{m}\frac{d\Psi_{g}^{*}}{dz}\right)dS\label{eq:matrix}
\end{equation}

are the matrix elements for the transition, with $m_{0}$ the electron mass in the dipole layer. The terms $\Psi_{g}(\mathbf{r},z)$ and $\Psi_{m}(\mathbf{r},z)$ represent the graphene and metal electron wavefunctions, respectively. Then, inserting Eq. (\ref{eq:gamma}) into Eq. (\ref{eq:current}), the tunneling current can be expressed as

\begin{equation}
I=g_{V}\frac{4\pi e}{\hbar}\sum_{g,m}|M_{gm}|^{2}\left[f_{g}(E_{g})-f_{m}(E_{m})\right]\delta(E_{g}-E_{m}).
\label{eq:current2}
\end{equation}

Considering the graphene with two identical atoms per unit cell, labeled 1 and 2, the wavefunction for wavevector $\mathbf{k}$ can be written in terms of the basis functions $\Phi_{j\mathbf{k}}(j=1,2)$ on each atom as $\Psi_{g}(\mathbf{r},z)=\chi_{1}(\mathbf{k}_{g})\Phi_{1\mathbf{k}_{g}}(\mathbf{r},z)+\chi_{2}(\mathbf{k}_{g})\Phi_{2\mathbf{k}_{g}}(\mathbf{r},z)$. The basis functions have Bloch form, $\Phi_{j\mathbf{k}_{g}}(\mathbf{r},z)=exp\left(i\mathbf{k}_{g}\cdot\mathbf{r}\right)u_{j\mathbf{k}_{g}}\left(\mathbf{r},z\right)/\sqrt{A}$, where $u_{j\mathbf{k}_{g}}(\mathbf{r},z)$ is a periodic function and $A$ refers to the contact area. These periodic functions are localized around the basis atoms (i.e., as $2p_{z}$ orbitals) of the graphene , and $u_{j\mathbf{k}_{g}}\left(\mathbf{r},z\right)$ is expected to vary only weakly along the radial coordinate $\mathbf{r}$ in the graphene. Thus, we assume that $u_{j\mathbf{k}_{g}}(\mathbf{r},z)=f_{j\mathbf{k}_{g}}\left(\mathbf{r}\right)g(z)$ and we approximate the radially-dependent term $f_{j\mathbf{k}_{g}}\left(\mathbf{r}\right)$ as numerical constants $f_{1}$ and $f_{2}$ \cite{Feenstra}. The $z$-dependence has the usual decaying form $g(z)=\sqrt\kappa e^{-\kappa z}$, where
$\kappa$ is the decay constant of the wavefunction in the barrier. The decay constant $\kappa$ has the form $\sqrt{2m\phi/\hbar^{2}+k_{\parallel}^{2}}$ \cite{Feenstra}, where $\phi\sim W_m$ is the barrier height in the dipolar layer and $k_{\parallel}$ is the parallel momentum. For graphene, the latter term is essentially equal to the momentum at the K or K' points (i.e., $4\pi/3a)$ so that $\kappa\sim20$nm$^{-1}$ for $W_m\sim5$eV. 

Both $\chi_{1}(\mathbf{k}_{g})$ and $\chi_{2}(\mathbf{k}_{g})$ have well-known values for graphene in a nearest-neighbor tight-binding approximation \cite{key-RMP}, 

\begin{equation}
\left[\begin{array}{c}
\chi_{1}\\
\chi_{2}
\end{array}\right]=\frac{1}{\sqrt{2}}\left[\begin{array}{c}
e^{\mp i\alpha/2}\\
se^{\pm i\alpha/2}
\end{array}\right]\label{eq:8}
\end{equation}

where $\alpha$ is the angle of the relative wavevector, the upper sign is for the band extreme at the
K point of the Brillouin zone and the lower sign is for the K' point, with $s=+1$ for the conduction band (CB) and -1 for the valence band (VB). On the other hand, the metal electrons can be modeled as free incident and reflected particles for $z\geq d$ and with a decaying exponential for $z<d$, namely 

\begin{equation}
\Psi_{m}(\mathbf{r},z)=\begin{cases}
\begin{array}{c}
\frac{e^{i\mathbf{k}_{m}\cdot\mathbf{r}}}{\sqrt{V}}te^{\kappa(z-d)}\\
\frac{e^{i\mathbf{k}_{m}\cdot\mathbf{r}}}{\sqrt{V}}\left[e^{-ik(z-d)}+re^{ik(z-d)}\right]
\end{array} & \begin{array}{c}
z<d\\
z\geq d
\end{array}\end{cases}\label{eq:9}
\end{equation}

where $t$ and $r$ are the amplitudes of the transmitted and reflected waves, respectively. As usual, the matching conditions $\Psi_{m}\left(\mathbf{r},z\right)\mid_{z=d^{-}}=\Psi(\mathbf{r},z)|_{z=d^{+}}$ and $m_0^{-1}\left(d\Psi_{m}/dz\right)|_{z=d^{-}}=m^{-1}\left(d\Psi_{m}/dz\right)|_{z=d^{+}}$ have to be fulfilled, resulting in $t=2k_{z}/\left(k_{z}+i\kappa m/m_0\right)$. Thus, the matrix elements for the transitions of Eq. \ref{eq:matrix} can be written as

\begin{equation}
M_{gm}\approx\frac{\hbar^{2}}{2m_{0}}\frac{4k_{z}\kappa}{k_{z}+i\kappa \dfrac{m}{m_0}}\Theta\left(\alpha\right)\frac{e^{-\kappa d}}{\sqrt{L\cdot D}}\frac{1}{A}\int dSe^{i(\mathbf{k}_{g}-\mathbf{k}_{m})\cdot\mathbf{r}},\label{eq:10}
\end{equation}

where we have defined $\Theta\left(\alpha\right)=\chi_{1}^{\ast}f_{1}^{\ast}+\chi_{2}^{\ast}f_{2}^{\ast}$. The integral on the right-hand side of Eq. \ref{eq:10} approaches the delta-function $\delta(\mathbf{k}_{g}-\mathbf{k}_{m})$ when $A\rightarrow\infty$, implying the conservation of in-plane momentum $\mathbf{k}$: $\mid M_{gm}\mid^{2}\varpropto\mid A^{-1}\int dSe^{i(\mathbf{k}_{g}-\mathbf{k}_{m})\cdot\mathbf{r}}\mid^{2}\rightarrow\delta_{\mathbf{k}_{g},\mathbf{k}_{m}}^{2}=\delta_{\mathbf{k}_{g},\mathbf{k}_{m}}$. Incorporating Eq. \ref{eq:10} into Eq. \ref{eq:current2}, we get the following expression for the current

\begin{equation}
I=\frac{8\pi e}{\hbar}\left(\frac{\hbar^{2}}{2m_{0}}{4\kappa\sqrt\kappa e^{-\kappa d}}\right)^{2}\frac{1}{L}\sum_{\mathbf{k}_{g},\mathbf{k}_{m},k_{z}}\mid\Theta(\alpha)\mid^{2}\frac{k_{z}^{2}}{k_{z}^{2}+\left( \dfrac{m}{m_0}\kappa\right) ^{2}}\left[f_{g}(E_{g})-f_{m}(E_{m})\right]\delta\left(E_{g}-E_{m}\right)\delta_{\mathbf{k}_{g},\mathbf{k}_{m}}.\label{eq:11}
\end{equation}

\begin{figure}
\includegraphics[scale=0.5]{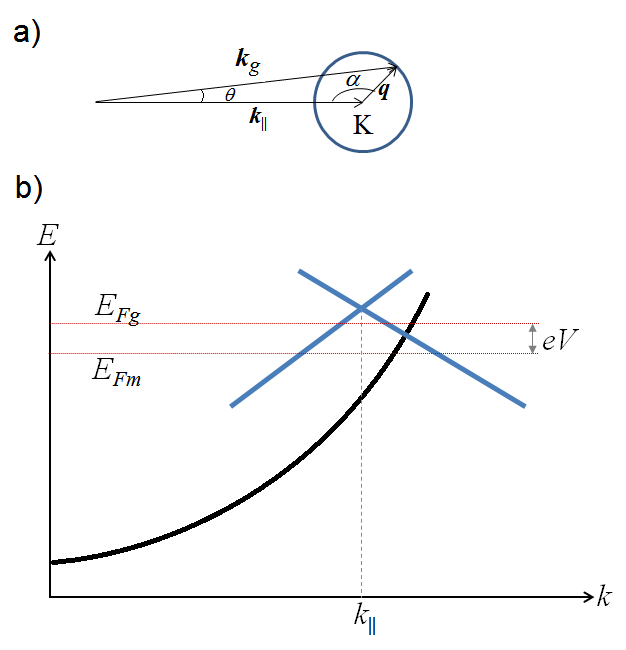}
\caption{a) Diagram of the graphene momentum relative $q$ to the K point.
b) Metal and graphene dispersion relations.}
\label{fig:reldis}
\end{figure}

The delta Dirac function guarantees that only energy-conserving tunneling
processes are possible. From the Fig. (\ref{fig:reldis}a) we observe that $\mathbf{k}_{g}=\mathbf{k}_{\parallel}+\mathbf{q}$,
with $\mathbf{k}_{\parallel}$ constant and thus $\sum_{\mathbf{k}_{g}}\Leftrightarrow\text{\ensuremath{\sum}}_{\mathbf{q}}.$
The function $\mid\Theta(\alpha)\mid^{2}$ is
$\mid f_{1}\mid^{2}+sf_{1}^{\ast}f_{2}^{\ast}\cos\left(\alpha\right)$,
where $\mid f_{1}\mid^{2}$ is a constant of order unity assumed to
have no dependence on $k_g$.

In deriving Eq. \ref{eq:11} we have incorporated both the graphene
and metal dispersion relations, namely $E_{g}=E_{g}(\mathbf{k}_{g})=E_{D}\pm\hbar v_{f}q$
and $E_{m}=E_{m}(\mathbf{k}_{m},k_{z})=\hbar^{2}(k_{m}^{2}-k_{z}^{2})/2m$,
which we have sketched in Fig. (\ref{fig:reldis}b) for convenience
and $k_{g}^{2}=k_{\parallel}^{2}+q^{2}-2qk_{\parallel}\cos\alpha$. Considering
Eq. \ref{eq:11} in the limit of large $A$, $\mathbf{k}_{g}=\mathbf{k}_{m}=\mathbf{k}$,
and then the equation for the tunneling current becomes

\begin{equation}
I=\frac{8\pi e}{\hbar}\left(\frac{\hbar^{2}}{2m_{0}}\frac{4\kappa e^{-\kappa d}}{\sqrt{D}}\right)^{2}\frac{1}{L}\sum_{\mathbf{q,}k_{z}}\mid\Theta(\alpha)\mid^{2}\omega\left(k_{z}\right)\left[f_{g}(E_{g})-f_{m}(E_{m})\right]\delta\left(E_{g}-E_{m}\right),\label{eq:12}
\end{equation}
where we have defined the function $\omega\left(k_{z}\right)=k_{z}^{2}/\left(k_{z}^{2}+(m\kappa/m_0)^{2}\right)$.
The discrete sums over $\mathbf{q}$ and $k_{z}$ are converted to
integrals using the recipes $\sum_{\mathbf{q}}\rightarrow A/(2\pi)^{2}\iint d\alpha qdq$
and $\sum_{k_{z}}\rightarrow L/2\pi\int dk_{z}$.
After some algebra, the tunnel current density becomes

\begin{equation}
J=\eta\left(\kappa\right)\iiint d\alpha dqdk_{z}q\mid\Theta(\alpha)\mid^{2}\omega\left(k_{z}\right)\left[f_{g}\left(E_{g}\right)-f_{m}\left(E_{m}\right)\right]\delta\left(E_{g}-E_{m}\right),\label{eq:13}
\end{equation}

where $\eta\left(\kappa\right)=\frac{16e}{\hbar}\left(\frac{\hbar^{2}}{2m_{0}}\frac{\kappa\sqrt{\kappa} e^{-\kappa d}}{\pi}\right)^{2}$. The energy difference appearing in the delta-function can be written as,

\begin{equation}
E_{g}-E_{m}=\frac{\hbar^{2}}{2m}\left(k_{z}^{*2}-k_{z}^{2}\right)=0,\label{eq:14}
\end{equation}

where $k_z^{*2}=q^2+2q\left(\xi-k_{\parallel}\cos\alpha\right)+k_\parallel^2-k_D^{2}$ with $\xi=mv_f/\hbar$,$k_D^2=2m/\hbar^2E_D$.
  Using the Dirac delta function properties we can write $\delta\left[ \frac{\hbar^2}{2m}\left(k_z^{*2}-k_z^{2}\right) \right] =2m/\hbar^2\delta\left(k_z-k_z^{*}\right)/\left|k_z+k_{z}^{*}\right|$ and Eq. \ref{eq:13} becomes 

\begin{equation}
J=\eta\left(\kappa\right)\frac{2m}{\hbar^{2}}\iint d\alpha dqq\mid\Theta(\alpha)\mid^{2}\frac{k_{z}^{*}}{2\left(k_{z}^{*2}+(m\kappa/m_0)^{2}\right)}eV\frac{\partial f}{\partial\mu}.\label{eq:13-1}
\end{equation}

Since we are interested in the specific contact resistivity (i.e $V\rightarrow0$) we have approximated the Fermi levels difference by $eV\frac{\partial f}{\partial\mu}$, where $\mu=(E_{Fg}+E_{Fm})/2$. Given that $q\ll k_{\parallel}$ is fulfilled, we can approximate $k_{z}^{*2}\approx k_{\parallel}^{2}-k_{D}^{2}$ to find an analytical solution for Eq (\ref{eq:13-1}). Thus Eq. (\ref{eq:13-1}) can be expressed as

\begin{equation}
J=\eta\left(\kappa\right)\frac{2m}{\hbar^{2}}\iint d\alpha dqq\mid\Theta(\alpha)\mid^{2}\frac{\sqrt{k_{\parallel}^{2}-k_{D}^{2}}}{2\left(k_{\parallel}^{2}-k_{D}^{2}+(m\kappa/m_0)^{2}\right)}eV\frac{\partial f}{\partial\mu}.
\end{equation}

Now integrals of the type

\begin{equation}
\intop_{0}^{\infty}q\frac{\partial f}{\partial\mu}dq=\intop_{0}^{\infty}q\frac{\exp\left[\left(E-\mu\right)/kT\right]}{kT\left(1+\exp\left[\left(E-\mu\right)/kT\right]\right)^{2}}dq,
\end{equation}
where $E-\mu=eV/2+\Delta E-\hbar v_{f}q$, have to be resolved for
every cone. Thus, the current density takes the form

\begin{equation}
J=\eta\left(\kappa\right)\frac{2m}{\hbar^{2}}\frac{\pi eV}{\left(\hbar v_{f}\right)^{2}}\ln\left(\exp\left[\left(eV/2+\Delta E\right)/kT\right]+1\right).
\end{equation}
Finally, an analytical expresssion for the specific contact resistivity
$\rho_{c}=\left(dJ/dV\right)^{-1}\mid_{V=0}$ as the Eq. (5)
in the main text is deduced. 
\subsection{Effect of the broadening on the contact resistance}
The two components of the contact resistance must be modified to take into account the broadening of the states in both graphene under the metal and graphene in the channel, namely;

\begin{equation}
R_{mg}^{-1}(\Delta E_m)=\int \tilde R_{mg}^{-1}(E_1)G_1(E_1-\Delta E_m;t_1)dE_1,
\label{eq:Rmg_b}
\end{equation}

\begin{equation}
R_{gg}^{-1}(\Delta E_m,\Delta E_g)=\iint \tilde R_{gg}^{-1}(E_1,E_2)G_1(E_1-\Delta E_m;t_1)G_2(E_2-\Delta E_g;t_2)dE_1dE_2,
\label{eq:Rgg_b}
\end{equation}

where $\tilde R_{mg}$ and $\tilde R_{gg}$ are given by Eqs. (4) and (9) of the main text and the Gauss function is $G(x-x_0;t)={exp[(x-x_0)^2/t^2]}/(t\sqrt\pi)$.
\subsection{Ni-Graphene junction}
\begin{figure}
\begin{subfigure}[t]{.47\textwidth}
  \caption{}
  \includegraphics[scale=0.260]{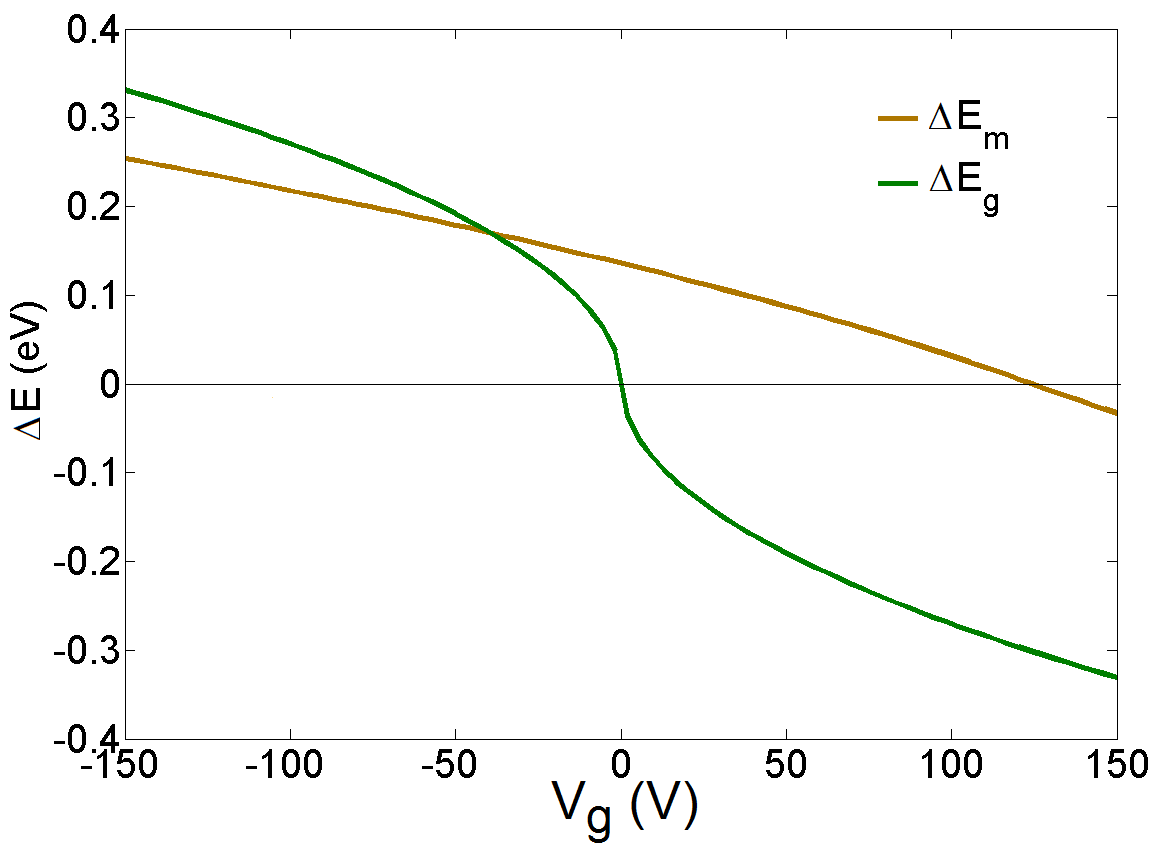}
  \label{fig:DeltaE_Nia}
\end{subfigure}
\quad
\begin{subfigure}[t]{.47\textwidth}
  \caption{}
  \includegraphics[scale=0.25]{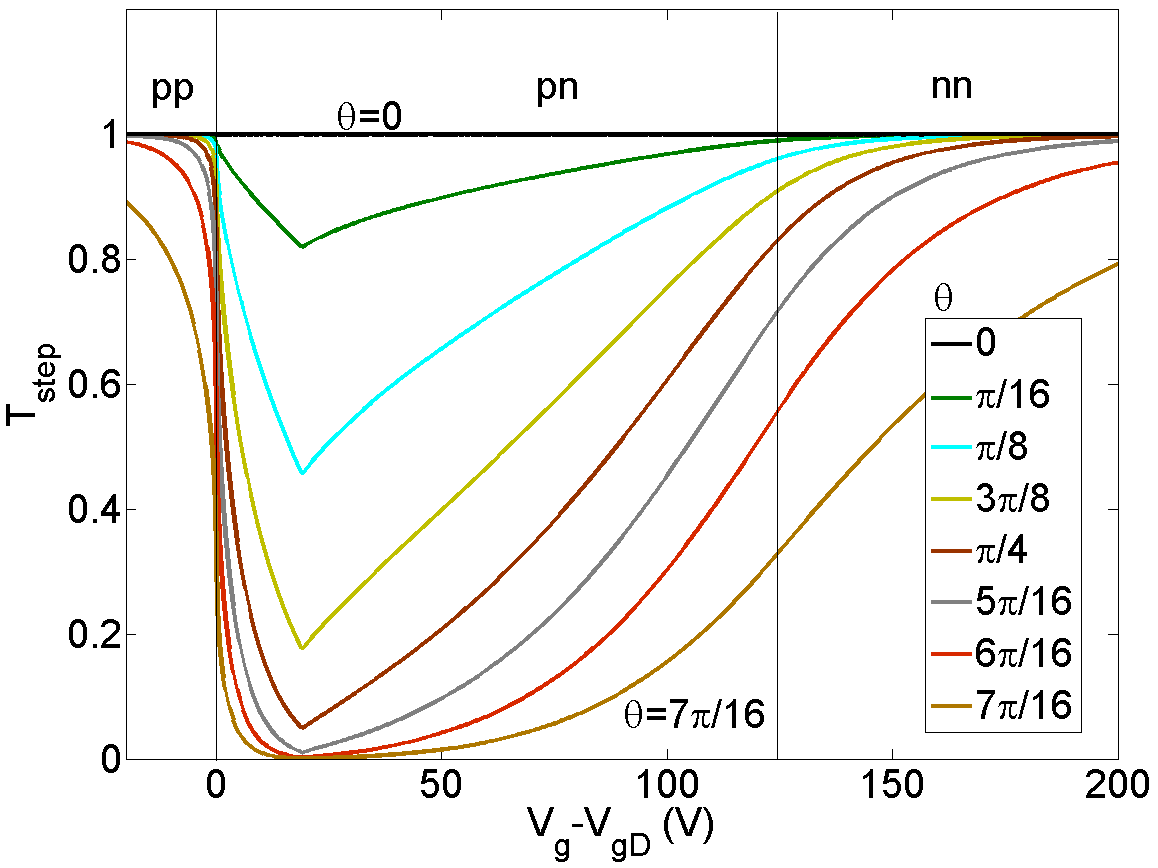}
  \label{fig:DeltaE_Nib}
\end{subfigure}
\caption{Electrostatic behavior and transmission of the Ni-graphene contact: (a) Graphene Fermi level shifts under the contact and in the channel with respect to the Dirac point as a function of the back-gate voltage $V_{g}$. (b) Transmission probability through a potential step with $\lambda=4$nm for different angles of incidence. Electrical parameters for this metal were taken from Table I of the main text.}
\label{fig:DeltaE_Ni}
\end{figure}

Fig. \ref{fig:DeltaE_Nia} shows the shift of the Fermi level respect the Dirac point for the Ni-graphene junction. Important values are $V_{gD}\sim0$V and $V_D\sim125$V defining the crossover between unipolar pp-junction/bipolar pn-junction and bipolar pn-junction/unipolar nn-junction, respectively. The electrical parameters for this simulation have been mentioned in the main text. On the other hand, Fig. \ref{fig:DeltaE_Nib} shows the transmission probability of the Dirac fermions across the potential step for different incidence angles.

\subsection{Ti-Graphene junction}
\begin{figure}
\begin{subfigure}[t]{.47\textwidth}
  \caption{}
  \includegraphics[scale=0.260]{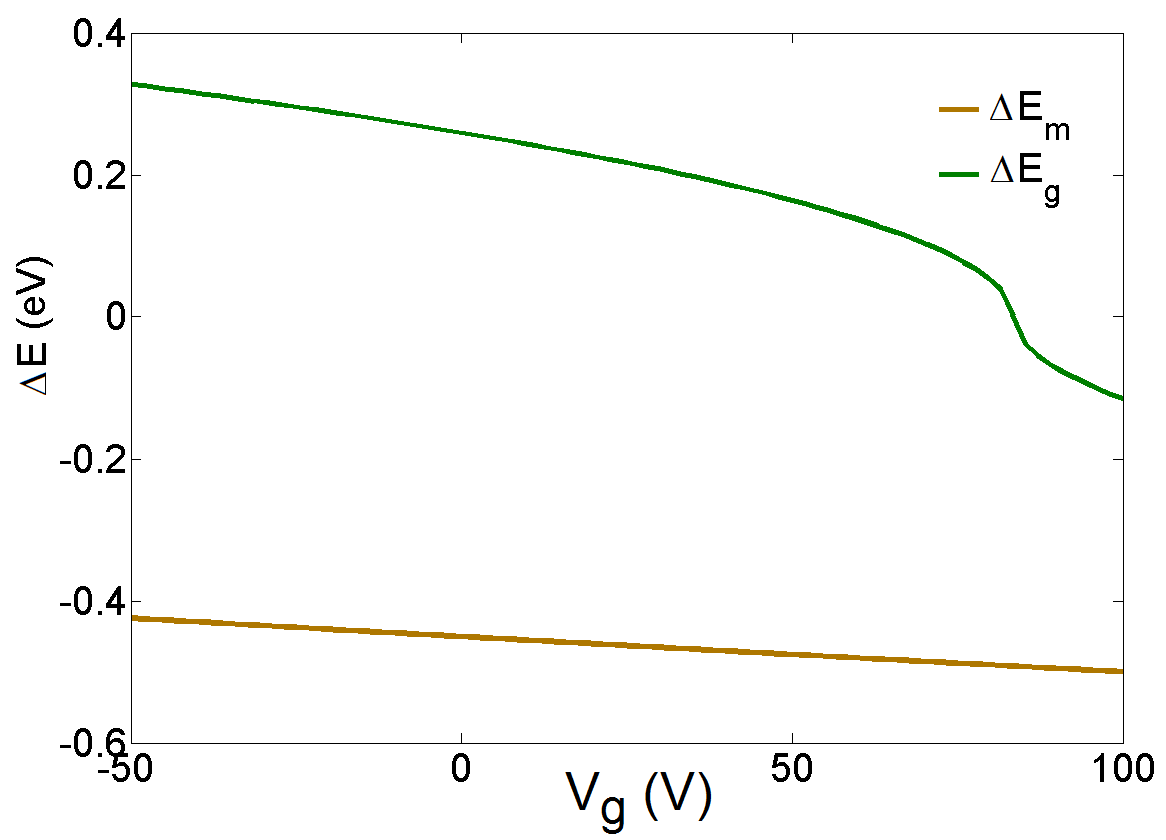}
  \label{fig:DeltaE_Tia}
\end{subfigure}
\quad
\begin{subfigure}[t]{.47\textwidth}
  \caption{}
  \includegraphics[scale=0.25]{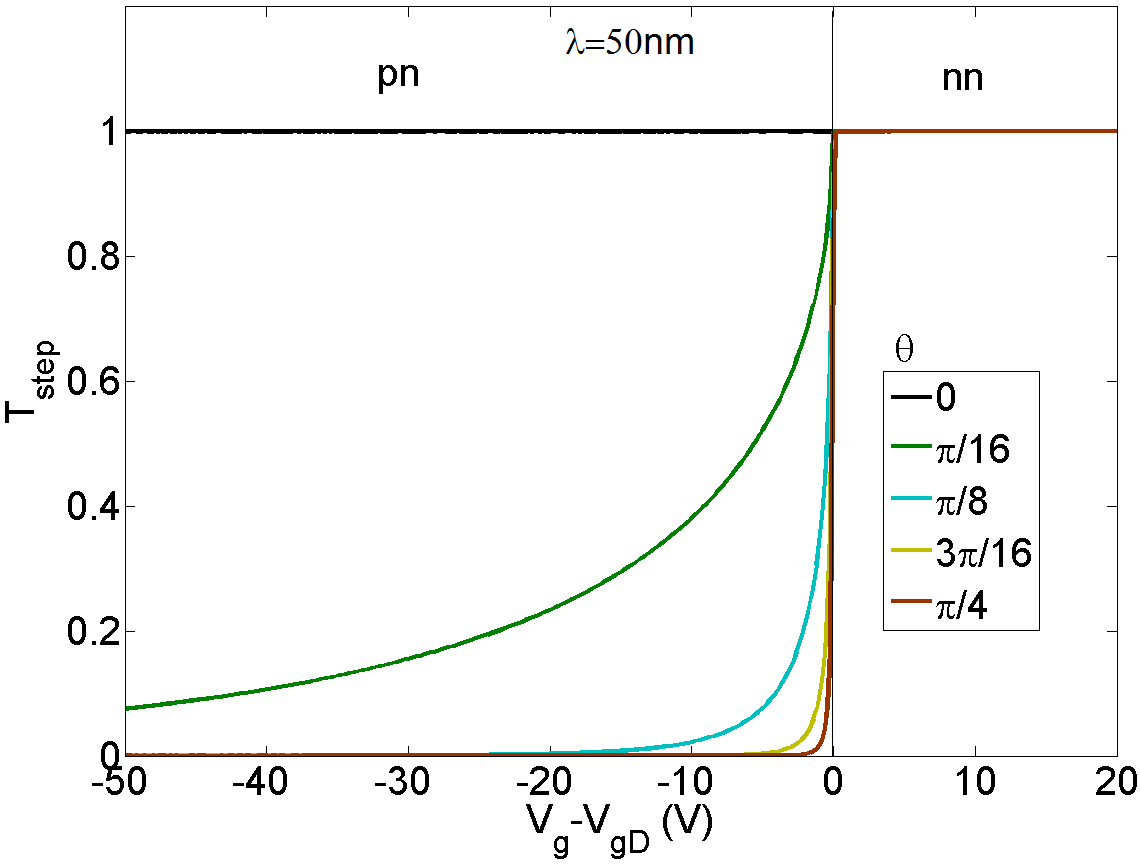}
  \label{fig:DeltaE_Tib}
\end{subfigure}
\caption{Electrostatic behavior and transmission of the Ti-graphene contact: (a) Graphene Fermi level shifts under the contact and in the channel with respect to the Dirac point as a function of the back-gate voltage $V_{g}$. (b) Transmission probability through a potential step with $\lambda=50$nm for different angles of incidence. Electrical parameters for this metal were taken from Table I of the main text.}
\label{fig:DeltaE_Ti}
\end{figure}

Fig. \ref{fig:DeltaE_Tia} shows the shift of the Fermi level respect the Dirac point for the Ti-graphene junction. Important value here is $V_{gD}\sim75$V defining the crossover between bipolar pn-junction/unipolar nn-junction. The electrical parameters for this simulation have been mentioned in the main text.  On the other hand, the Fig. \ref{fig:DeltaE_Tib} shows the transmission probability of the Dirac fermions across the potential step for different incidence angles.

\end{document}